\documentstyle[aps,preprint]{revtex}

\newcommand{\diag}{\,\mbox{\rm diag}\,}

\newcommand{\bbeta}{\mbox{\boldmath $\beta$\unboldmath}}
\begin{document}

\title{Hartle-Hawking state in supersymmetric minisuperspace}
\author{Andr\'as Csord\'as\thanks{Permanent address: Research Institute
for Solid State Physics, P.O. Box 49, H1525 Budapest, Hungary} and
Robert Graham}
\address{Fachbereich Physik, Universit\"at-Gesamthochschule Essen\\
45117 Essen\\ Germany}
\maketitle

\begin{abstract}
The Hartle-Hawking `no-boundary' state is constructed explicitely for the
recently developed supersymmetric minisuperspace model with non-vanishing
fermion number.
\end{abstract}

Spatially homogeneous models both in gravity and in supergravity have
enjoyed some popularity in recent years as a testing ground for new ideas
in quantum cosmology. One such idea, which has been discussed extensively
in the literature, is the proposal by Hartle and Hawking for the
construction of the `wave-function of the universe', including gravity
\cite{HH}.
According to this proposal the quantum state of the universe is formally
given by the Euclidean path-integral of exp[-action] over all compact
4-geometries, containing a given compact 3-geometry (the argument of the
wave-function) as its only boundary. This is why it is also called the
`no-boundary' state.
While this idea of striking (but also deceptive) simplicity could
be partially implemented, e.g. in spatially homogeneous minisuperspace
models, like a closed Friedmann universe with a scalar field \cite{HH}
or an anisotropic Bianchi type IX universe with a cosmological
constant \cite{Louko} its use in {\it supersymmetric} minisuperspace
models has caused some difficulty.

The supersymmetric Friedmann model without matter was treated successfully
 \cite{3} but lacks sufficient degrees of freedom to permit a physically
meaningful discussion of this issue. The inclusion of a spatially
homogeneous  supersymmetric scalar matter-field has, so far, led only
to explicit solutions of the wormhole-type \cite{3,Death}.
The first treatments of the spatially homogenous supersymmetric anisotropic
Bianchi type IX model without matter field concluded that a
Hartle-Hawking state would not exist in such a model \cite{5,6,6a,6b},
the only permitted state being that of a `worm-hole' in the completely
empty or filled fermion sectors which had previously been found in
\cite{7}. Subsequently
it was shown \cite{8} that the particular $SO(3)$ symmetry of Bianchi
type IX permits an alternative homogeneity ansatz for the
Rarita-Schwinger field, and that its application replaces the permitted
`worm-hole' state in the empty or filled fermion sector by a
`no-boundary' state in the same sector. In a recent paper \cite{9}
we reexamined the supersymmetric minisuperspace models of Bianchi type
in class A \cite{10} without matter fields
and showed that, contrary to previous expectations, they posses
{\it infinitely many} physical states. Hence, the question of the
existence and form of a `no-boundary' state in such models must be
reconsidered. In the present paper we (i) apply the theory of \cite{9}
to the
supersymmetric Bianchi type IX model without matter, and with the
conventional homogeneity condition for the Rarita-Schwinger field,
and (ii) construct the Hartle-Hawking `no-boundary'-state for that
model explicitely. The dependence of that state on the 3-metric turns out
to be the same as in \cite{8} (see also \cite{11}), where the alternative
homogeneity condition was applied. However, the dependence on the
spatially homogeneous Rarita-Schwinger field is completely different
from \cite{8}.
It turns out to be a state near the middle of the fermion number spectrum,
between the completely empty and the completely filled fermion sectors.
This state has
a much better chance to permit an extension to full supergravity, because
it was proven that the physical states in full supergravity cannot ly in
the empty and filled fermion sectors \cite{13}.
A brief account of our results has already been given in a recent
conference report \cite{Csordas}.

Let us begin recapitulating some notation and results of \cite{9}
which are necessary here. The starting point is the Langrangean of
$N=1$ supergravity in the notations defined in \cite{14}.
Space-time is assumed to be foliated by space-like 3-surfaces which
are homogeneous under the action of a 3-dimensional homogeneity group
which is here assumed to be $SO(3)$. A symmetric basis of 1-forms
$\omega^p$ then exists $(p=1,2,3)$, satisfying $h^{1/2}d\omega^p=
\frac{1}{2}\delta^{pq}\epsilon_{qrs}\omega^r\wedge\omega^s$, where
$h_{pq}$ with $h=\det h_{pq}$ are the purely time-dependent components
of the spatial 3-metric, and $\epsilon_{qrs}$ are the components of the
3-dimensional Levi-Civita tensor. The volume of the underlying
3-sphere is $V=\int\omega^1\wedge\omega^2\wedge\omega^3=16\pi^2$. In
the metric representation the independent variables are given by the
spatial components of the tetrad ${e_p}^a(a=0,1,2,3)$ satisfying
${e_p}^ae_{qa}=h_{pq}$, and the spatial components of the Grassmannian
Rarita-Schwinger field ${\psi_p}^\alpha$,
${\bar{\psi}_p}^{\ \dot{\alpha}}$. We shall here adopt the homogeneity
conditions ${e_p}^a={e_p}^a(t)$, ${\psi_p}^\alpha={\psi_p}^\alpha(t)$
and shall {\it not} make use of the alternative homogeneity condition
for ${\psi_p}^\alpha$ consistent with $SO(3)$ which was proposed in
\cite{8}. Introducing canonical momenta, Poisson brackets and finally
Dirac brackets in order to eliminate the appearing second class
constraints one finds canonical expressions for the supersymmetry
generators $S_\alpha$, $\bar{S}_{\dot{\alpha}}$ and the Lorentz generators
$J_{\alpha\beta}$, $\bar{J}_{\dot{\alpha}\dot{\beta}}$
of the following form
\begin{eqnarray}
\label{eq:1}
  S_\alpha &=& - {\cal C}_{pr}^{\dot{\alpha}\beta}
            \left(\frac{1}{2}V\delta^{pq}{e_q}^a+{\textstyle\frac{i}{2}}
                 {p_+}^{pa}\right)
   \sigma_{a\alpha\dot{\alpha}}
  {\pi^r}_\beta\nonumber\\
  \bar{S}_{\dot{\alpha}} &=& \left(\frac{1}{2}V\delta^{pq}{e_q}^a
     -{\textstyle\frac{i}{2}}{p_+}^{pa}\right)
    \sigma_{a\alpha\dot{\alpha}}{\psi_p}^\alpha
\end{eqnarray}
and
\begin{eqnarray}
\label{eq:2}
  J_{\alpha\beta} &=& +{\textstyle\frac{1}{2}}
       (\sigma^{ac}\epsilon)_{\alpha\beta}
    \left(e_{pa}{{p_+}^p}_c-e_{pc}{{p_+}^p}_a\right)
  \nonumber\\
      && -{\textstyle\frac{1}{2}}
         \left(\psi_{p\alpha}{\pi^p}_\beta+\psi_{p\beta}
           {\pi^p}_\alpha\right)
  \nonumber\\
  \bar{J}_{\dot{\alpha}\dot{\beta}} &=& -{\textstyle\frac{1}{2}}
    (\epsilon\bar{\sigma}^{ac})_{\dot{\alpha}\dot{\beta}}
     \left(e_{pa}{{p_+}^p}_c-e_{pc}{{p_+}^p}_a\right)\,.
\end{eqnarray}
For all conventions regarding the $\sigma$-matrices and
$\epsilon_{\alpha\beta}$ we refer to \cite{14}. The kernel
${\cal C}_{pq}^{\dot{\alpha\alpha}}$ is defined as
\begin{equation}
\label{eq:3}
  {\cal C}_{pq}^{\dot{\alpha\alpha}} = -{\textstyle\frac{1}{2Vh^{1/2}}}
    \left(ih_{pq}n^a-\epsilon_{pqr}e^{ra}\right)
       \bar{\sigma}_a^{\hphantom{a}\dot{\alpha}\alpha}
\end{equation}
$n^a$ is the future oriented unit vector normal on the space-like
3-surfaces and its components are functions of the ${e_p}^a$. The
variables ${p_+^p}_a$ and the Grassmannian ${\pi^p}_\alpha$ are the
`Dirac-conjugates' of ${e_p}^a$ and ${\psi_p}^\alpha$ in the
sense that the only non-vanishing Dirac-brackets are
\begin{eqnarray}
\label{eq:4}
\{{e_p}^a, {{p_+}^q}_b\}^* &=& {\delta_p}^q{\delta_b}^a\nonumber\\
 \{{\psi_p}^\alpha, {\pi^q}_\beta\}^* &=&
   -{\delta_p}^q{\delta_\beta}^\alpha\,.
\end{eqnarray}
Canonical quantization is performed in the metric
$({e_p}^a,{\psi_p}^\alpha)$-representation by
putting
\begin{equation}
\label{eq:5}
    {{p_+}^p}_a=-i\hbar(\partial /\partial{e_p}^a) \,\qquad
   {\pi^p}_\alpha = -i\hbar(\partial /\partial{\psi_p}^\alpha)\,.
\end{equation}
 There is an ordering ambiguity in the expression for $S_\alpha$ because
the kernel (\ref{eq:3}) does not commute with ${p_+^p}_a$. Here we
shall deviate from ref. \cite{9} and adopt the choice of the ordering
as displayed explicitely in eq.~(\ref{eq:1}), while
in \cite{9} we ordered the kernel ${\cal C}_{pr}^{\dot{\alpha}\beta}$
to the right of ${p_+}^{pa}$ before quantizing. While, at least so far,
no reason of principle is visible to prefer one choice of ordering over
the other (or over any mixed ordering in between), the ordering
chosen here will actually simplify in an essential way the form of
eq.~(\ref{eq:22})
below.

With the adopted choice of operator ordering we find the explicit graded
generator algebra
\begin{eqnarray}
\label{eq:6}
\Big[S_\alpha,S_\beta\Big]_+&=&0=
 \Big[\bar{S}_{\dot{\alpha}},\bar{S}_{\dot{\beta}}\Big]_+\\
\label{eq:7}
  \Big[S_{\alpha},\bar{S}_{\dot{\alpha}}\Big]_+&=&
    - \frac{\hbar}{2}H_{\alpha\dot{\alpha}}\\
\label{eq:8}
  \Big[H_{\alpha\dot{\alpha}},S_{\beta}\Big]_- &=&
    -i\hbar\varepsilon_{\alpha\beta}
       {\bar{D}_{\dot{\alpha}}}^{\hphantom{p}{\dot{\beta}\dot{\gamma}}}
        \bar{J}_{\dot{\beta}\dot{\gamma}}\\
\label{eq:9}
   \Big[H_{\alpha\dot{\alpha}},\bar{S}_{\dot{\beta}}\Big]_- &=&
      i\hbar\varepsilon_{\dot{\alpha}\dot{\beta}}{J}_{\beta\gamma}
 {D_\alpha}^{\beta\gamma}\nonumber\\
 &=& i\hbar\varepsilon_{\dot{\alpha}\dot{\beta}}
     \Big[{D_\alpha}^{\beta\gamma}
{J}_{\beta\gamma}+i\hbar{\bar{E}_\alpha}^{\phantom{p}\dot{\gamma}
\dot{\delta}}
\bar{J}_{\dot{\gamma}\dot{\delta}}-\frac{i\hbar n^a}{Vh^{1/2}}
   \sigma_{a\alpha\dot{\gamma}}\bar{S}^{\dot{\gamma}}
\Big]
\end{eqnarray}
and the well-known commutators with $J_{\alpha\beta}$,
$\bar{J}_{\dot{\alpha}\dot{\beta}}$ reflecting Lorentz transformations.
The operator $H_{\alpha\dot{\alpha}}$ is here {\it defined} by the
anti-commutator (\ref{eq:7}), but we have checked that it
classically differs only by terms proportional to Lorentz generators
from $\tilde{H}_{\alpha\dot{\alpha}}$ defined by the diffeomorphism
and Hamiltonian generators $H^p$ and $H$ via
$\tilde{H}_{\alpha\dot{\alpha}}=\sigma_{a\alpha\dot{\alpha}}({e_p}^a
H^p+{n}^aH)$. The structure functions ${D_\alpha}^{\beta\gamma}$,
${\bar{D}_{\dot\alpha}}^{\dot{\beta}\dot{\gamma}}$,
${\bar{E}_\alpha}^{\dot{\gamma}\dot{\delta}}$ are Grassmannian
odd functions of ${e_p}^a$, ${\psi_p}^\alpha$. While their explicit
form is not essential, for the following, we shall here list them
for completeness and future reference
\begin{eqnarray}
\label{eq:10}
 {D_\alpha}^{\beta\gamma} &=& n^b{e_p}^c\varepsilon^{\beta\delta}
    (\sigma_b\bar{\sigma}_c)_\delta^{\ \gamma}
      \Bigg[h^{-1/2}\delta^{pq}\varepsilon_{\alpha\rho}{\psi_q}^{\rho}+
       \epsilon^{pqr}{\sigma_a}_{\alpha\dot{\alpha}}
        C_{sq}^{\dot{\alpha}\sigma}\varepsilon_{\sigma\rho}
        {\psi_r}^{\rho}\bigg(\frac{V}{2}\delta^{st}{e_t}^a+
         \frac{i}{2}{p_+}^{sa}\bigg)\Bigg]\\
\label{eq:11}
 {\bar{E}_\alpha}^{\ \dot{\gamma}\dot{\delta}} &=&
  \left(\frac{i}{2Vh^{1/2}}\right)\left(n^ae^{pb}+n^be^{pa}\right)
   \epsilon_{\alpha\gamma}{\bar{\sigma}_a}^{\ \dot{\gamma}\gamma}
    {\bar{\sigma}_b}^{\ \dot{\delta}\beta}\psi_{p\beta}
\end{eqnarray}

${\bar{D}_{\dot{\alpha}}}^{\dot{\beta}\dot{\gamma}}$ and
${E_{\dot{\alpha}}}^{\gamma\delta}$ are given by the
matrix-adjoints of these expressions. Due to the different ordering
chosen, the algebra (\ref{eq:6})-(\ref{eq:9}) differs slightly from
a corresponding result given in \cite{9}, but both forms are, of course,
fully consistent.

As all generators in (\ref{eq:6})-(\ref{eq:9}) appear on the right-hand
side the `graded' algebra closes not only classically, but also quantum
mechanically. Due to the Jacobi-identity for commutators this result is
even sufficient to prove that the only remaining commutator
$[H_{\alpha\dot{\alpha}}, H_{\beta\dot{\beta}}]_-$ evaluates to structure
functions multiplied with generators $S_\gamma$, $\bar{S}_{\dot{\gamma}}$,
$J_{\gamma\delta}$, $\bar{J}_{\dot{\gamma}\dot{\delta}}$,
$H_{\gamma\dot{\gamma}}$ on the right,
i.e. we find that this spatially homogeneous model has a closed generator
algebra and is free from anomalies.

Let us now turn to the physical states of the system in the sense of Dirac
\cite{Dirac}, i.e. the states which are annihilated by all the generators
$S_\alpha$, $\bar{S}_{\dot{\alpha}}$, $J_{\alpha\beta}$,
$\bar{J}_{\dot{\alpha}\dot{\beta}}$, $H_{\alpha\dot{\alpha}}$. These
states $\Psi_F$ can be parametrized by the conserved fermion number
${\psi_p}^\alpha\partial/\partial{\psi_p}^\alpha=F$ and have the
form
\begin{eqnarray}
\label{eq:12}
 \Psi_0 &=& \exp\left[\frac{V}{2\hbar}\delta^{pq}h_{pq}\right]\\
\label{eq:13}
 \Psi_2 &=& \bar{S}_{\dot{\alpha}}\bar{S}^{\dot{\alpha}} f(h_{pq})\\
\label{eq:14}
 \Psi_4 &=& S^\alpha S_\alpha g(h_{pq})\prod_{r=1}^3(\psi_r)^2\\
\label{eq:15}
 \Psi_6 &=& \exp\left[-\frac{V}{2\hbar}\delta^{pq}h_{pq}\right]
   \prod_{r=1}^3(\psi_r)^2\,.
\end{eqnarray}
Here the amplitudes $f$ and $g$ appearing in the 2- and 4-fermion sector,
respectively, are functions of the metric $h_{pq}$ only, which makes all
states (\ref{eq:12})-(\ref{eq:15}) Lorentz-invariant and serves to
satisfy the Lorentz-constraints. The functions $f$, $g$ satisfy
Wheeler-DeWitt equations, which are obtained by applying $S_\alpha$ to
$\Psi_2$ and $\bar{S}^{\dot{\alpha}}$ to $\Psi_4$, respectively, and
using the algebra (\ref{eq:6})-(\ref{eq:9}) \cite{7}. In the first case
we obtain
\begin{equation}
 \left(H_{\alpha\dot{\alpha}}^{(0)}-\frac{\hbar^2}{Vh^{1/2}}n^a
   \sigma_{a\alpha\dot{\alpha}}\right)f(h_{pq})=0
\label{eq:16}
\end{equation}
where we have used the identity $[\bar{S}^{\dot{\alpha}},
\sigma_{a\alpha\dot{\alpha}}n^a/h^{1/2}]=0$ to factor out
$\bar{S}^{\dot{\alpha}}$ to the left. We discarded the possibility that
the right hand side of eq.~(\ref{eq:16}) could be non-zero and
proportional to a bosonic function annihilated by
$\bar{S}^{\dot{\alpha}}$. The reason is that all such functions are known
to vanish in {\it full} supergravity \cite{13}. Here
$H_{\alpha\dot{\alpha}}^{(0)}$
consists only of the bosonic terms of $H_{\alpha\dot{\alpha}}$, i.e.
of those terms which remain if ${\pi^p}_\alpha$ is first brought to the
right using its anti-commutation relation with ${\psi_p}^\alpha$,
and is then equated to zero. In the 4-fermion sector we find in an
analogous manner
\begin{equation}
  H_{\alpha\dot{\alpha}}^{(1)}g(h_{pq})=0\,,
\label{eq:17}
\end{equation}
however, $H_{\alpha\dot{\alpha}}^{(1)}$ is now obtained from
$H_{\alpha\dot{\alpha}}$ by bringing ${\psi_p}^\alpha$ to the right, using
its anti-commutation relation with ${\pi^p}_\alpha$, and then equating
it to zero.

To get explicit expressions it is useful to parametrize the spatial
metric by $h_{pq}=\Omega_{pi}(e^{2\bbeta})_{ij}\Omega_{qj}$ where
$\Omega_{pi}$ is a rotation matrix, depending on three Euler angles,
and
\begin{equation}
\label{eq:18}
  \left( e^{2\bbeta}\right)_{ij}= e^{2\alpha}\diag
    \left(e^{2\beta_++2\sqrt{3}\beta_-},
    e^{2\beta_+-2\sqrt{3}\beta_-},
     e^{-4\beta_+}\right)\,.
\end{equation}
It is important to note that the
rotation matrix $\Omega_{pi}$ and the parameters $\alpha$, $\beta_+$,
$\beta_-$ are unique functions of the tetrad ${e_p}^a$.
The diffeomorphism constraint
\begin{equation}
  {e_p}^a{\bar{\sigma}_a}^{\dot{\alpha}\alpha}
  {H^{(0)}}_{\alpha\dot{\alpha}}f(h_{pq})=0=
   {e_p}^a\bar{\sigma}_a^{\dot{\alpha}\alpha}
    {H^{(1)}}_{\alpha\dot{\alpha}}g(h_{pq})
 \label{eq:19}
\end{equation}
is then satisfied by taking $f(h_{pq})$ and $g(h_{pq})$ as
{\it independent} of the Euler angles of the rotation matrices, thus
$f=f(\alpha,\beta_+,\beta_-)$, $g=g(\alpha,\beta_+,\beta_-)$. There only
remains the Hamiltonian constraint
\begin{eqnarray}
\label{eq:20}
 && n^a{\bar{\sigma}_a}^{\dot{\alpha}\alpha}
  \Bigg({H_{\alpha\dot{\alpha}}}^{(0)}-\frac{\hbar^2}{Vh^{1/2}}
   n^b\sigma_{b\alpha\dot{\alpha}}\Bigg)f(\alpha,\beta_+,\beta_-)=0\\
\label{eq:21}
 && n^a{\bar{\sigma}_a}^{\dot{\alpha}\alpha}{H_{\alpha\dot{\alpha}}}^{(1)}
    g(\alpha,\beta_+,\beta_-)=0\,.
\end{eqnarray}
The latter reads explicitely,
\begin{eqnarray}
\label{eq:22}
&&  \Bigg[
     -\frac{\hbar^2}{V^2}\left(\frac{\partial}{\partial\alpha}\right)^2+
\frac{\hbar^2}{V^2}
       \left(\frac{\partial}{\partial\beta_+}\right)^2+\frac{\hbar^2}{V^2}
        \left(\frac{\partial}{\partial\beta_-}\right)^2+
         \left(\frac{\partial\phi}{\partial\alpha}\right)^2-
          \left(\frac{\partial\phi}{\partial\beta_+}\right)^2-
           \left(\frac{\partial\phi}{\partial\beta_-}\right)^2\nonumber\\
&& \mbox{\hspace{2.5cm}}
         +\frac{\hbar}{V}\left(-\frac{\partial^2\phi}{\partial\alpha^2}+
             \frac{\partial^2\phi}{\partial\beta_+^2}+
              \frac{\partial^2\phi}{\partial\beta_-^2} \right)\Bigg]
               g(\alpha,\beta_+,\beta_-)=0
\end{eqnarray}
with the abbreviation
\begin{equation}
\label{eq:23}
 \phi=\frac{1}{2}\delta^{pq}h_{pq}=
    \frac{1}{2}e^{2\alpha}\left(2e^{2\beta_+}\cosh2\sqrt{3}
\beta_-+e^{-4\beta_+}\right)\,.
\end{equation}
Due to our judicious choice of ordering in eq.~(\ref{eq:1}) a term
proportional to $\frac{\hbar^2}{V}e^{-3\alpha}g(\alpha,\beta_+,\beta_-)$
is avoided in eq.~(\ref{eq:21}), while the corresponding term is present
in eq.~(\ref{eq:20}), which we will not need in the following, however.
In fact we shall here only be interested in some special explicit
solutions of eq.~(\ref{eq:22}), as it will turn out that the
Hartle-Hawking state we are looking for is among them. First we note
that a very simple solution of eq.~(\ref{eq:22}) is given by
$g(h_{pq})\sim\exp(-V\phi(h_{pq})/\hbar)$, but this solution, inserted
in eq.~(\ref{eq:14}), gives $\Psi_4=0$, i.e. it only gives the trivial
solution. Remarkably, however, there are four equally simple linearly
independent further
solutions of eq.~(\ref{eq:22}) which give nontrivial results for $\Psi_4$.
The first of these is the desired Hartle-Hawking state, namely
\begin{eqnarray}
\label{eq:24}
&& g(h_{pq})=\exp\Big[-\frac{V}{2\hbar}e^{2\alpha}\big(2e^{2\beta_+}
     (\cosh 2\sqrt{3}\beta_--1)\nonumber\\&&  \mbox{\hspace{4.3cm}}+
   e^{-4\beta_+}-4e^{-\beta_+}\cosh\sqrt{3}\beta_-\big)\Big]\,.
\end{eqnarray}
The other three states are
 \begin{eqnarray}
\label{eq:24a}
&& g(h_{pq})=\exp\Big[-\frac{V}{2\hbar}e^{2\alpha}\big(4e^{2\beta_+}
     (\sinh(\sqrt{3}\beta_-)^2+e^{-4\beta_+}\nonumber\\
&&  \mbox{\hspace{4.3cm}}
   +4e^{-4\beta_+}\cosh\sqrt{3}\beta_-\big)\Big]\,.
\end{eqnarray}
and the two further expressions obtained by rotating the
$(\beta_+,\beta_-)$-axis around $\beta_+=0=\beta_-$ twice by
120$^\circ$-degrees, respectively. The final form of the Hartle-Hawking
state in the 4-fermion sector, i.e. $\Psi_4$, is obtained as a function
of ${\psi_p}^a$ and ${e_p}^a$ by acting with the operator
$(S^\alpha S_\alpha)$ on $g(h_{pq})\Pi(\psi_r)^2$. To perform this step
one should express the invariants $\alpha$, $\beta_+$, $\beta_-$ of
the spatial metric in terms of the matrix elements $h_{pq}$ which are
functions of the tetrad via the relation $h_{pq}={e_p}^ae_{qa}$.

Let us now discuss the result further. The result (\ref{eq:24})
coincides in form with the amplitude of the
Hartle-Hawking state in the filled-fermion sector found in \cite{8}
by assuming a different homogeneity condition for the Rarita-Schwinger
field. By contrast, here we have assumed the usual homogeneity condition
${\psi_p}^\alpha={\psi_p}^\alpha(t)$ and the amplitude (\ref{eq:24}),
via eq.~(\ref{eq:17}), corresponds to a state in the middle of the
fermion-number spectrum, namely in the 4-fermion sector. This change
is highly wellcome, because quantum states in the empty and filled
fermion sector are known \cite{13} not to exist in full
supergravity, where the existing states are grouped around the middle
of the fermion-number spectrum, corresponding to the Dirac-vacuum of the
gravitino. Therefore the state (\ref{eq:14}), (\ref{eq:24}) may now well
have a counterpart in full supergravity.

That eq.~(\ref{eq:24}) indeed
gives the Hartle-Hawking state can be seen as follows: First of all
the real exponential form of $g(h_{pq})$ shows that no classically allowed
domain of the spatial metric is described by this wave-function. (This
is an agreement with the known fact that no empty closed Friedmann
universe can exist classically, but may exist as a quantum fluctuation.
However, it is in contrast to the classical possibility of an empty
anisotropic Bianchi-type IX mixmaster universe \cite{16}. Such
classically allowed mixmaster solutions must therefore correspond to
{\it other} solutions of eqs.~(\ref{eq:16}) or (\ref{eq:17})). The spatial
metric therefore exists in this wave-function only due to classically
forbidden tunnelling processes. To exhibit these in a semi-classical
way let us write $g(h_{pq})$ in the form
$g(h_{pq})\sim\exp[-\frac{V}{\hbar}{\mbox{\rm I}}(h_{pq})]$ thereby
defining the Euclidean action
${\mbox{\rm I}}={\mbox{\rm I}}(\alpha,\beta_+,\beta_-)$. Then the
semi-classical(i.e. most probable) tunnelling path parametrized by
a suitable affine parameter $\lambda$ satisfies the first-order
differential equations
\begin{eqnarray}
\label{eq:25}
p_\alpha &=& \frac{\partial{\mbox{\rm I}}}{\partial\alpha}=
  -\frac{d\alpha}{d\lambda}\nonumber\\
p_{\beta_\pm} &=& \frac{\partial{\mbox{\rm I}}}{\partial\beta_\pm}=
  \frac{d\beta_\pm}{d\lambda}\,.
\end{eqnarray}
With solutions $\alpha(\lambda)$, $\beta_+(\lambda)$, $\beta_-(\lambda)$
the corresponding 4-metric has the form
\begin{equation}
\label{eq:26}
 ds^2=\left(3\sqrt{V}e^{3\alpha}d\lambda^2+(e^{2\bbeta})_{pq}
   \omega^p\omega^q\right)\,.
\end{equation}
Eqs.~(\ref{eq:26}) with ${\mbox{\rm I}}=\frac{1}{2}e^{2\alpha}
(2e^{2\beta_+}(\cosh2\sqrt{3}\beta_--1)+e^{-4\beta_+}-4e^{-\beta_+}
\cosh\sqrt{3}\beta_-)$
are, in fact, well known \cite{17}. They have been solved \cite{17}
to give the 4-metric of a compact Riemannian 4-space filling in,
without singularity, any given
3-geometry of Bianchi type IX whose metric tensor is parametrized by
$\alpha,\beta_-,\beta_+$. For our spatially homogeneous model this is
the property which defines the `no-boundary' state, at least
semi-classically.But since eq.~(\ref{eq:24}) also solves the fully
quantum mechanical Wheeler DeWitt equation (\ref{eq:21}) is an exact
quantum
amplitude with the required semiclassical property and hence, indeed
the exact Hartle-Hawking state of the supersymmetric Bianchi-type IX model.
The states (\ref{eq:24a}) can be discussed in a
similar manner. However, in these cases, the semiclassical tunnelling
paths extending the given 3-geometry turn out to describe
{\it non-compact} 4-geometries, as one of the scale-parameters grows
without bound in the limit $\alpha\rightarrow-\infty$, even though the
other two scale-parameters and the metric 3-volume shrink to zero.
Hence, these states (and similarly the states $\Psi_0$, $\Psi_6$ of
(\ref{eq:12}), (\ref{eq:15})) do not qualify as `no-boundary states'.

In summary, giving an explicit solution of all constraints of a quantized
supersymmetric spatially homogenous cosmological model without matter or
cosmological constant we have found a state in one of the sectors in the
middle of the spectrum of fermion numbers which qualifies as the
`no-boundary' state of this system. The explicit form (\ref{eq:24}) shows
that this state, for values of the overall scale-parameter $e^\alpha$
large compared to the Planck-length, strongly favors isotropic metrics
$(\beta_+,\beta_-\to 0)$. It will, of course, be interesting to extend
this analysis e.g. by allowing for a cosmological constant \cite{18},
or a matter field, or treating the case of full supergravity \cite{19}.
While such
extensions are technically more demanding the present analysis gives
clear indications how one may proceed.

This work has been supported by the Deutsche Forschungsgemeinschaft through
the Sonderforschungsbereich 237 ``Unordnung und gro{\ss}e Fluktuationen''.
One of us (A.~Csord\'as) would like to acknowledge additional support by
The Hungarian National Scientific Research Foundation under Grant number
F4472.

\end{document}